\documentstyle[prl,aps,floats,twocolumn,epsfig]{revtex}
\begin{document}
\twocolumn[\hsize\textwidth\columnwidth
\hsize\csname @twocolumnfalse\endcsname

\title
{Directed polymers with tilted columnar disorder
and Burgers-like turbulence}
\author
{Roya Mohayaee$^{1,2}$, 
Attilio L. Stella$^{1}$
and Carlo Vanderzande$^{3,4}$}
\address
{$^1$
INFM-Dipartimento di Fisica and Sezione INFN, Universit\`{a} di Padova, 
I-35131 Padova, Italy\\
$^2$ Dipartimento di Fisica, Universita degli Studi di Roma ``La 
Sapienza'', I-100185 Roma, Italy\\
$^3$ Departement WNI, Limburgs Universitair Centrum, 
3590 Diepenbeek, Belgium\\
$^4$ Instituut voor Theoretische Fysica, Katholieke Universiteit Leuven,
3001 Heverlee, Belgium\\}

\maketitle

\begin{abstract}

The minimal energy 
variations of a directed polymer with tilted columnar disorder in
two dimensions are shown numerically to obey a multiscaling at 
short distances which crosses
over to global simple scaling at large distances. The 
scenario is analogous to that of structure functions in bifractal
Burgers'
turbulence.
Some scaling properties are predicted 
from extreme value statistics. The multiscaling disappears
for zero tilt.
\\
PACS numbers: 05.40.-a, 47.53.+n, 74.60.Ge, 02.50.-r

\end{abstract}
\hspace{.2in}
]

The directed polymer in a random medium (DPRM) is a
prototype model in the statistical physics of disordered
systems \cite{HHR}. At zero temperature, $T$, for each realisation
of a random potential landscape,
the problem consists in the statistical characterisation
of long directed paths (no overhangs)
which minimise the total energy.
Besides having ramifications and generalizations in different
fields, the DPRM is directly related to flux lines in
type II superconductors, domain walls in random ferromagnets
and nonequilibrium fluctuations of growing 
interfaces in dimension $d=2$
\cite{barabasibook}.
In the last case the DPRM energies can be mapped onto
the interfacial profiles realised during the
growth, and the quenched average over potential disorder
amounts to a mean over growth histories.

Attention has been devoted in recent years to the effects
of various forms of disorder on the asymptotic
DPRM scaling properties. Particular interest
focused on long-range disorder correlations which
in the language of interfacial growth, can be either in 
"time" $t$ (longitudinal spatial variable) or 
in space $x$ (transversal) \cite{medina},
\cite{lam3}.
While the effects of spatial correlations are relatively
well understood, long range temporal correlations 
(including those appropriate to $1/f$ noise) remain more 
problematic mainly due to a lack of Galilean invariance.
Here, we study the case of a noise which
is infinitely correlated along one space-time direction
($x=-t$), a situation we
refer to as tilted columnar disorder \cite{HHR,kruglast}.
In the DPRM context, columnar disorder is of particular
interest for describing flux lines in superconductors
irradiated by heavy ions \cite{HHR}. 
The case in which the columns are tilted with
respect to the (time) direction of the
flux lines could be of relevance, e.g., for
the behaviour of a flux line in the presence of a
forest of splayed columnar defects. Such
situations have been studied recently both theoretically
and experimentally \cite{fscdt,fscde}.

In this Letter we show that the DPRM with tilted
columnar disorder in $d=2$ and at $T=0$ is characterised
by two distinct critical regimes. The first is a global
one, holding at large length scales 
and consistent with simple scaling.
We fully elucidate this regime in terms of
extreme value statistics. The second regime is local and holds
at smaller length scales. It is characterised by 
intermittency and multifractal scaling. The two regimes match
in a full scaling framework which displays close analogies and
correspondences with the scaling regimes of
a fully turbulent flow \cite{frisch,lohse,krug}. 
The very existence of this unexpected
local multiscaling is especially 
remarkable because the distributions
generating it are quite standard and general, and do not
necessarily involve long tails. 

Consider a directed polymer that evolves
on a $1+1$ dimensional lattice
in a random energy landscape $\varepsilon(x,t)$.
The energies $\varepsilon(x,t)$ associated to each site
$(x,t)$ are taken from a probability distribution 
function (pdf) $P(\varepsilon)$.
The partition function $Z(x,t)$ is then defined through
the recursion relation
\begin{eqnarray}
Z(x,t) = e^{-\varepsilon (x,t)/T} [ Z(x-1,t-1)+Z(x+1,t-1)]
\label{1}
\end{eqnarray}
together with an initial condition (we take
$Z(x,t=0)=e^{-\varepsilon(x,0)/T}$ with
$\varepsilon(x,0)$ randomly distributed according to $P(\varepsilon)$).
We will in particular study tilted columnar defects, 
$\varepsilon(x,t)=\varepsilon(x+t)$, and will compare them to the more standard
columnar defects, $\varepsilon(x,t)=\varepsilon(x)$
\cite{kruglast}. Fig. 1 sketches
the situation for the tilted case.

\begin{figure}
  \centerline{ \epsfig{figure=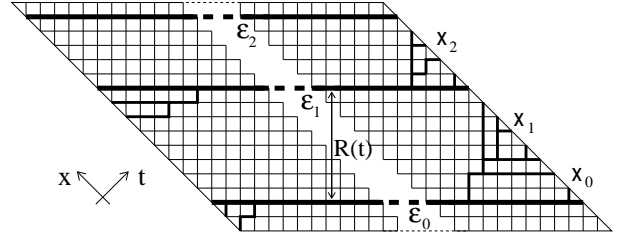,width=8cm}}
\vskip 0.5cm
\caption{
Tilted columnar defects are oriented along lines $x+t$ constant. 
At $T=0$, the
optimal paths (thin lines) follow columns with very low energies
$\varepsilon_0,\varepsilon_1,\varepsilon_2,...$ (thick lines) for
a very large fraction of time (see text). Periodic boundary conditions
are applied at the horizontal sides of the strip.}
\label{figure 1}
\end{figure}

It is generally assumed \cite{HHR} that the recursion (\ref{1}) is
a discrete version of the diffusion equation 
\begin{eqnarray}
\frac{\partial Z}{\partial t} = \frac{T}{2} \frac{\partial^2 Z}{\partial x^2} -
\frac{1}{T}
V(x,t) Z
\label{2}
\end{eqnarray} 
where $V(x,t)$ is a random
potential with
zero mean. 
From (\ref{2}) one obtains
 the KPZ \cite{kpz} equation for the height $h$ of a growing interface
by the usual Hopf-Cole transformation $h=T \log{Z}$
\begin{eqnarray}
\frac{\partial h}{\partial t}=\frac{T}{2}\frac{\partial^2 h}{\partial x^2}
+ \frac{1}{2} \left(\frac{\partial h}{\partial x}\right)^2 - V(x,t)
\label{2b}
\end{eqnarray}
Finally, (\ref{2b}) can in turn be mapped onto the Burgers equation
\cite{burgers}
\begin{eqnarray}
\frac{\partial u}{\partial t} + u \frac{\partial u}{\partial x}=
\frac{T}{2} \frac{\partial^2 u}{\partial x^2}
 + \eta(x,t)
\label{2a}
\end{eqnarray}
by letting $u=-\partial h/\partial x$,
where $u$ is interpreted as the velocity of a fluid
subject to a random force $\eta(x,t)=-\partial V/\partial x$,
and with viscosity $T/2$.

Here, we will limit ourselves to the $T \to 0$ limit of (\ref{1}).
Denoting $\lim_{T\to 0} T \log{Z}=-E$, (\ref{1}) reduces to
\begin{eqnarray}
E(x,t) = \min[E(x-1,t-1), E(x+1,t-1)]
+ \varepsilon(x,t)
\label{3}
\end{eqnarray}

Taking $\varepsilon=0$ in (\ref{3}) one obtains, in the
continuum limit, an equation with a nonlinearity 
$|\partial h/\partial x|$ \cite{KS1}. Noise disorder generated when $\varepsilon
\neq 0$ is however expected to transform such nonlinearity into the
analytical one of (\ref{2b}).
Using the numerical methods of ref. \cite{KS2}, we found evidence
of such analytical nonlinearity
\cite{voetnota}.


In Ref. \cite{kruglast}, some properties of the minimal energy path
in columnar disorder ($\varepsilon=\varepsilon(x)$) were studied numerically.
It was found that the disorder averaged transversal
displacement $R(t)=\langle x_{\mbox{min}}(t) \rangle$
of the minimal energy path
grows as a power of time $R(t) \sim t^{1/z}$, where $z$
is highly non-universal, and can be calculated for various
pdf's $P$ using Flory arguments based on
extreme value statistics.
In particular for a Gaussian pdf, $z=1$,
but with logarithmic corrections. These arguments
should remain valid for the tilted columnar disorder we study.
Indeed, $z$ remains unchanged as we
verified numerically for a few distributions
$P(\varepsilon)$. 

Here, we study the moments of the energy-energy
structure function
\begin{eqnarray}
G_q(r,t) = \langle | E(x+r,t) - E(x,t) |^q \rangle^{1/q}
\label{4}
\end{eqnarray}
and find that for the case of tilted defects,
and for a very large class of pdf's $P(\varepsilon)$,
$G_q(r,t)$ shows two scaling regimes.
The first one is global for $r > R(t) \sim t^{1/z}$, with
$G_q(r,t)$ independent of $r$ and with $q$-independent simple scaling
in time
\begin{eqnarray}
G_q(r,t) \sim R(t)^\alpha \sim t^{\alpha /z}
\ \ \ \ \ r>R(t)
\label{5}
\end{eqnarray}
In contrast, for $r < R(t)$, there is strong evidence of multiscaling
\begin{eqnarray}
G_q(r,t) \sim r^{\zeta_q} \ \ \ \ \  1<r< R(t), \mbox{ t fixed}
\label{6}
\end{eqnarray}
Figure 2 shows our numerical results for 
$P(\varepsilon)=5|\varepsilon|^{-6}$
 with $\varepsilon \leq -1$.
These results were obtained from a numerical iteration of (\ref{3})
on a strip with periodic
boundary conditions and a typical width $2^{15}$
in both space and time directions. 
The results were averaged over
$2\times 10^3$ realisations of the random energy landscape.
Qualitatively similar behaviour was found for a Gaussian, and for power law 
pdf's like $P_+(\varepsilon)= \mu
\epsilon^{-\mu-1} \ (\varepsilon \geq 1)$ or even for 
pdf's with a bounded support like $P_\nu(\varepsilon)=(\nu+1)\varepsilon^{\nu}, \varepsilon \in [0,1]$.

\begin{figure}
  \centerline{ \epsfig{figure=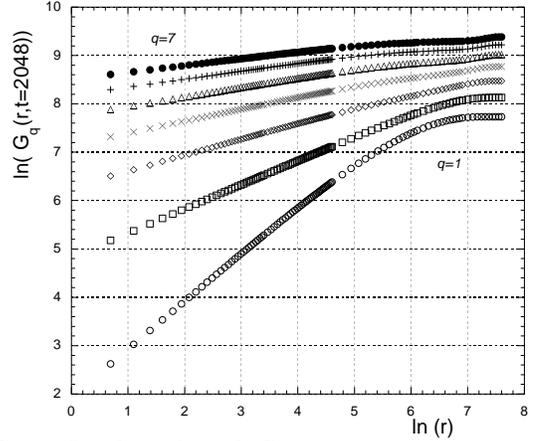,width=7cm}}

\caption{Log-log plot of $G_{q}(r,t=2048)$ versus $r$ for
$q=1,\ldots,7$. The data are for $P(\varepsilon)=
5|\varepsilon|^{-6}$ with $\varepsilon \leq -1$}

\label{figure 2}
\end{figure}

We also investigated in detail the nearest neighbour energy difference
which, when replaced by a space derivative, should correspond 
to the local velocity $u$ in (\ref{2a}).
Also this quantity shows multiscaling, i.e.
\begin{eqnarray}
\langle |u|^q\rangle^{1/q} (t) \simeq G_q(r=1,t) \sim t^{\alpha_q /z}
\label{7}
\end{eqnarray}
where $\alpha_q$ is $q$-dependent.
In Figure 3 we show our results for 
$\langle |u|^q\rangle^{1/q} (t)$ for a Gaussian distribution.
Qualitatively similar behaviour was found
for all $P$'s investigated.

\begin{figure}
  \centerline{ \epsfig{figure=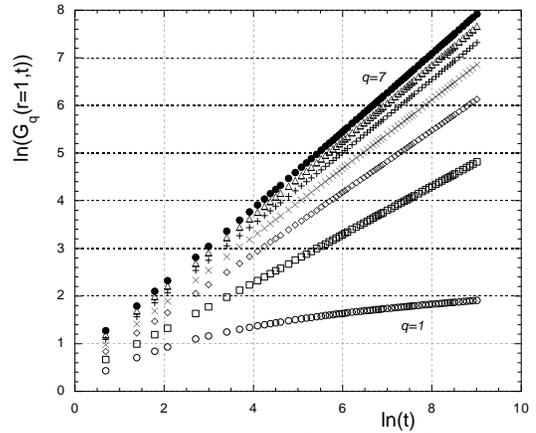,width=7cm}}
\caption{Log-log plot of $G_{q}(r=1,t)$ versus $t$ for 
$q=1,\ldots,7$.
The data are for a standard normal distribution.}

\label{figure 3}
\end{figure}

The scaling behaviour of $G_q(r,t)$ as given in (\ref{5}),(\ref{6})
and (\ref{7}) can in fact be described by a form which was originally
proposed for the structure functions of 
fully developed turbulence \cite{lohse}, and later
also for height correlators of certain models of epitaxial growth \cite{krug}:
\begin{eqnarray}
G_q(r,t) \sim R(t)^{\alpha_q} r^{\zeta_q} f_q\left(\frac{r}{R(t)}\right)
\label{8}
\end{eqnarray}
where $f_q(u) \to \mbox{constant}$ for $u \to 0$, and $f_q(u)
\to u^{-\zeta_q}$ when $u \to \infty$. Consistency with (\ref{5})
then implies the scaling identity
\begin{eqnarray}
\alpha_q + \zeta_q = \alpha
\label{9}
\end{eqnarray}
In the scalings (\ref{4})-(\ref{9}) the DPRM minimal energy $E(x,t)$ 
plays the role of a fluid velocity component in turbulence,
while $R$ corresponds to the Reynolds number.
Moreover, $[G_2(r=1,t)]^2$ is an analog of the local
average energy dissipation \cite{frisch,lohse,krug}.
In all cases investigated, (\ref{9}) holds within the
numerical accuracy. As an example, Table 1 shows our estimates
for $\alpha_q, \zeta_q$ and $\alpha$ for some $q$-values for the
case of $P_2(\varepsilon)=3\varepsilon^{2}$ with $\varepsilon \in [0,1]$. 
Since in all cases $q \zeta_q \sim 1$ for all $q>1$, while roughly
$q \zeta_q \sim q$ for $q<1$, the multiscaling appears
consistent with a biscaling, with $\zeta_q$ very
universal with respect to the noise distribution. 
On the other hand, as discussed below,
$\alpha$ is less universal, and its classes
appear to be at least as many as the possible asymptotic distributions
in extremal value statistics. 
Biscaling is a form of multiscaling already detected for
velocities in Burgers turbulence \cite{bouchaud}.
We will give full details of all exponent
estimates for different $P$'s elsewhere.
Eq. (\ref{8}-\ref{9}) reveal that, in presence of a tilt,
the variations of
minimal energy corresponding to even relatively
small $r$ are wildly fluctuating in the same way as
longitudinal velocity differences in a turbulent flow.
\begin{table}
\centerline{Table 1: Numerical estimates of exponents for 
$P_2(\varepsilon)=3\varepsilon^{2}$}
\centerline{with $\varepsilon \in [0,1]$. The error in these exponents
is estimated}
\centerline{to be a few percent.}
\begin{tabular}{l|lllllll}
q & .2 & .6 & 1 & 2 & 3 & 4 & 5 \\
\hline
$\zeta_{q}$ & 1.05 & 1.05 & .97 & .5 & .33 & .24 & .19 \\
$\alpha_{q}$ & .04 & 0.01 & .09 & .55 & .71 & .80 & .84 \\
$\alpha$   & 1.03  & 1.04 & 1.04 & 1.04 & 1.04  & 1.05  & 1.05 \\
\end{tabular}
\label{tabel 1}
\end{table}

The geometry of minimal
energy paths suggests an analytic determination of $\alpha$.
Consider the minimal energy path starting somewhere
and arriving in $x_0$ at time $t$ (Fig. 1). Typically, this path will
have followed a column with a favourable low energy $\varepsilon_0$ for a 'macroscopic'
fraction of $t$. The particular path chosen will depend on a balance
between the energy gained there and the energy it costs to get from
the starting point to the column of energy $\varepsilon_0$ and then to $x_0$ in
time $t$. As time goes on it pays to 'travel' to more distant but
energetically more favourable columns. Now consider paths arriving 
at $x_1 \neq x_0$. For small $|x_0-x_1|$, these paths will in general have
followed for a large fraction of $t$ the same low energy column. Thus,
the 
difference in energy between paths ending at $x_0$ and at $x_1$
will arise in a short part of their histories, where they
did not follow the same column. 
Considering sites $x_2$ that are further and further away from
$x_0$, we will arrive at the situation where it is more favourable
for the path ending at $x_2$ to follow another low energy column with
energy $\varepsilon_1$, closer to $x_2$, for most of its history
(Fig. 1). The typical distance between the different
minimal energy columns must be of the order of $R(t)$, the only 
relevant length scale
in the problem.
For the most part, each path
follows the column with the lowest energy within a local
region of size $R(t)$. Subsequently, at scales above
$R(t)$ the energies of different paths are essentially
independent, which accounts for the constant behaviour
of $G_q$ on these scales visible in Fig. 2.
As explained above, the
columns that attract most of the paths in a region of size $R$
are such that their energy is minimal in a region of that
size. The minimum $\varepsilon$ in a set of $R$ 
independent random variables
drawn from $P(\varepsilon)$
is distributed according to 
\begin{eqnarray}
\tilde{P}_R(\varepsilon) = R P(\varepsilon) \left[1- \int_{-\infty}^
{\varepsilon} P(\varepsilon') d\varepsilon' \right]^{R-1}
\label{10}
\end{eqnarray}
Since for $r>R(t)$ energy differences between
two paths are, up to small corrections,
given by the energy differences of the minimal energy columns they follow,
and since these are expected to be independent, we may write
\begin{eqnarray}
[G_q(r,t)]^q \approx t^q \int \int |\varepsilon_1 - \varepsilon_2|^q
\tilde{P}_R(\varepsilon_1) \tilde{P}_R(\varepsilon_2) d\varepsilon_1 d\varepsilon_2
\label{11}
\end{eqnarray}
It is not possible to evaluate (\ref{11})
for arbitrary pdf's analytically. However, from
extreme values statistics \cite{embrechts} it is known that for $R \to \infty$,
$\tilde{P}_R$ converges
to a limited number of possible asymptotic forms which only depend on the behaviour of
$P(\varepsilon)$ at $-\infty$.
As an example, for a Gaussian $P$,
or for other pdf's that approach zero at $-
\infty$ faster then any power,
$\tilde{P}_R(\varepsilon)$ converges to the Gumbel pdf
$P_G(y)=\exp(y-\exp y)$ where 
$y=\sqrt{2 \log{R}}(\varepsilon + \sqrt{2 \log{R}})$.
Inserting this into (\ref{11}) one finds
\begin{eqnarray}
[G_q(r,t)]^q \sim t^q (\log{R})^{-q/2}
\label{12}
\end{eqnarray}
Since  for the Gaussian
distribution $R(t) \sim t/\log{t}$ \cite{kruglast},
$G_q(r,t)$ indeed satisfies (\ref{5}), with $\alpha=1$,
up to logarithmic corrections.
One can proceed similarly for other pdf's.
For distributions whose support is
bounded from below such as $P_+$, or $P_\nu(\varepsilon) $,
 $\tilde{P}_{R}$ converges for $R \to \infty$ to 
the
Weibull distribution, from which the universal
value $\alpha=1$ is again obtained, but without logarithmic corrections.
Finally, for pdf's approaching zero as a power law
at $-\infty$ , i.e. $P_{-}(\varepsilon) \sim |\varepsilon|^{-\mu-1}$, $\alpha$ is 
non universal and equals $1+1/\mu$.
These predictions are consistent with our estimates
$\alpha \sim 1.03$ ($P_{+}(\varepsilon)$ with $\mu=3$),
$1.05$ ($P_{2}(\varepsilon)$) and $1.19$ ( 
$P_{-}(\varepsilon)$ with $\mu=5$). For the Gaussian case
the data can be fitted well by the form (\ref{12}).

We verified that when the tilt is removed and when
$\varepsilon(x,t)=\varepsilon(x)$ is assumed, the multiscaling
disappears for all $P$'s studied. However, 
$\alpha$ remains unchanged. Indeed, our
arguments for the its determination
should be valid both in presence and in absence of tilt.

The difference between these two cases can be understood
by referring again to Fig.1. Let us assume that
$\varepsilon_1$ is particularly low. 
Only paths whose endpoint is above that column can take advantage
of this low energy. As a consequence, there is a very big energy
difference for optimal paths ending at sites $x$ that are close to, but
on different sides of the column with energy $\varepsilon_1$.
In contrast, if the tilt is absent, two optimal paths ending on different sides
of a very low energy column 
can both follow that column for a long
time, and hence energy differences tend to be smaller.
The large energy differences in the tilted case
lead to broad distributions and multiscaling.

It remains an interesting question to find out whether the multiscaling
observed here
is present for any value of the tilt angle, or whether it only
appears above a certain non-zero critical angle.

In \cite{medina}, the KPZ equation 
(\ref{2b}) with noise $V(x,t)$ that has long range correlations
in time
\begin{eqnarray}
\langle V(x,t) V(x',t')\rangle \sim |t-t'|^{2\theta-1}\delta(x-x')
\label{14}
\end{eqnarray}
was studied. From a renomalisation group analysis it was expected
that for $\theta$ sufficiently below $1/2$
\begin{eqnarray}
(1+2\theta)z=2\alpha .
\label{15}
\end{eqnarray}
The case $\theta=1/2$ marks the transition to a
regime where infrared divergences become
important and the renormalisation procedure breaks down. 
Dimensionally, the noise corresponding to columnar
disorder in absence of tilt, shares the same
correlation (\ref{14}) with the $1/f$ noise ($\theta=1/2$). Thus, once
accepted that (\ref{2b}) at $T=0$ is the appropriate
continuum limit, our
results suggest that for pdf's (e.g. Gaussian) whose extreme value
statistics is given by the Gumbel distribution, (\ref{15})
remains valid at $\theta=1/2$, but with logarithmic corrections.

In summary, the differences in optimal energy
of the DPRM with tilted columnar defects obey a form of 
multiscaling analogous
to that valid for velocity structure functions in turbulence.
An intermittent scaling regime matches a global
simple scaling one. The global regime can be
characterized on the basis
of extreme value statistics arguments.  
The multiscaling at
small scales is an unexpected feature of both deterministic
and stochastic models of nonlinear diffusion. The multiscaling
of the discrete counterpart
of the velocity potential observed here
is similar to the scaling behaviour
of velocity increments of the Burgers equation solutions with 
other types of random forcing 
\cite{bouchaud}. 
The case treated here
is new also because, if (\ref{2a}) indeed provides
the appropriate continuum limit, the biscaling is valid
for increments of the velocity potential. 

The research presented here can be extended in several directions.
Besides obvious generalisations to finite temperature and higher
dimension, the case of networks of splayed columnar defects
deserves special attention \cite{fscdt}. It would be worth
investigating whether the multiscaling observed
here also shows up in these experimentally relevant situations.


{\bf Acknowledgement} 
We thank J. R. Banavar, J. Bec and U. Frisch for helpful discussions and M. Kardar 
for useful suggestions. Support from European Network Contract 
ERBFMRXCT980183, Italian MURST-cofin99 and Belgian IUAP
is gratefully acknowledged.


\begin{thebibliography}{99}
\bibitem{voetnota}
To establish more rigourously the validity of (\ref{2b})
and (\ref{2a}) at $T=0$ as continuum limits of (\ref{3}) for the
cases of present interest, is beyond our scope.
Here, we use these equations and the quantities they describe
mostly for comparison purposes.
\end{thebibliography}
\end{document}